# Electronic structure investigation of CoO by means of soft X-ray scattering

M. Magnuson, S. M. Butorin, J.-H. Guo and J. Nordgren

*Department of Physics, Uppsala University, P. O. Box 530, S-751 21 Uppsala, Sweden*

**Abstract**

The electronic structure of CoO is studied by resonant inelastic soft X-ray scattering spectroscopy using photon energies across the Co $2p$ absorption edges. The different spectral contributions from the energy-loss structures are identified as Raman scattering due to $d-d$ and charge-transfer excitations. For excitation energies close to the $L_3$ resonance, the spectral features are dominated by quartet-quartet and quartet-doublet transitions of the $3d^7$ configuration. At excitation energies corresponding to the satellites in the Co $2p$ X-ray absorption spectrum of CoO, the emission features are instead dominated by charge-transfer transitions to the $3d^8\underline{L}^{-1}$ final state. The spectra are interpreted and discussed with the support of simulations within the single impurity Anderson model with full multiplet effects which are found to yield consistent spectral functions to the experimental data.

## 1 Introduction

Electronic structures of transition metal compounds have been extensively investigated due to the discovery of exotic properties such as high-$T_c$ superconductivity, giant magnetoresistance and insulating behavior. For the theoretical description in ordinary band theory, the understanding of the electronic structure of transition metal oxides represents a fundamental problem since the metal ions show a more or less atomic-like behavior [1]. Band theory predicts that partially filled $3d$-electron compounds such as divalent CoO should be metallic opposite to their rather large band gaps determined by optical absorption spectroscopy (OAS) and their antiferromagnetic ionic insulator properties [2]. In contrast to the free ion, the degeneracy of the $3d$ states is partially lifted so that the $t_{2g}$ and $e_g$ states of the Co$^{2+}$ ion are energetically separated by the $O_h$ crystal-field splitting arising from the six octahedrally surrounding O$^{2-}$ ligands. The ground state of CoO posseses a spin quartet $^4T_{1g}(t_{2g})^5(e_g)^2$ symmetry and the hybridization of the metal ion states with the O $2p$ ligand states result in substantial charge transfer character. In the localized description of the $3d$ electrons, which is known to describe excitation spectra rather well, the $3d^7$, $3d^8\underline{L}^{-1}$ and $3d^9\underline{L}^{-2}$ configurations ($\underline{L}^{-1}$ denotes a hole in the O $2p$ ligand band) are energetically split up by the ligand field [3]. This implies several spin-quartet and doublet excited states within $\sim 3$ eV from the ground state. Due to ionic-lattice vibrations, direct $d-d$ transitions which are normally dipole forbidden become weakly allowed and have been identified in OAS [4,5]. These faint excitations can also be studied with electron energy loss spectroscopy (EELS) since the dipole selection rules are relaxed at low electron energies although these measurements are very surface sensitive [6, 7, 8]. While detailed resonant photoemission spectroscopy (RPES) measurements aimed at the $2p$ and $3p$ core levels [9,10,11,12,13] confirm the charge-transfer character of CoO, the spectra are mixed with the O $2p$ band and it is not clear to which extent the $3d^6$ and the screened $3d^7\underline{L}^{-1}$ final states reflects the electronic structure of the ground state [2]. Although these materials are known to exhibit charge-transfer insulator properties, this important mechanism is not yet fully understood.

In this work we investigate the electronic structure of CoO using resonant inelastic X-ray scattering (RIXS) spectroscopy with selective excitation energies around the Co 2$p$ thresholds. This technique is more bulk sensitive than RPES and each atomic element can be probed separately by tuning the excitation energy to the appropriate core edge. The $d-d$ excitations to the various excited states which can be studied in terms of loss structures become fully allowed due to the core-hole assisted excitation-deexcitation dipole transitions [14]. The 2$p$ spin-orbit coupling also allows Hund's rule electron (super)exchange scattering of quartet-doublet spin-flip transitions. When the excitation energy is tuned to the different features in the absorption spectrum, the RIXS spectra of CoO are found to exhibit resonant energy loss structures due to both $d-d$ excitations of Raman scattering and charge-transfer excitations of $3d^8\underline{L}^{-1}$ final state character. The RIXS spectra are interpreted with the support of multiplet calculations using the same set of parameters as in X-ray absorption. Although the final states of RIXS are slightly different from other spectroscopical techniques, it is useful to compare the energy positions of the peaks and the validity of multiplet calculations for the interpretation of the spectra.

## 2 Experimental Details

The measurements were performed at beamline BW3 at HASYLAB, Hamburg, using a modified SX700 monochromator [15]. The typical photon flux at the sample was about $2\times10^{12}$ photons/sec in 0.1% band width at 800 eV for 100 mA electron current in the storage ring. An XAS spectrum at the Co 2$p$ edges was obtained in total electron yield (TEY) by measuring the sample drain current. The Co $L_{2,3}$ RIXS spectra were recorded using a high-resolution grazing-incidence grating spectrometer with a two-dimensional position-sensitive detector [16]. During the XAS and RIXS measurements at the Co 2$p$ edges, the resolutions of the beamline monochromator were about 0.3 eV and 0.5 eV, respectively. The RIXS spectra were recorded with a spectrometer resolution better than 0.5 eV.

The measurements at the Co 2$p$ thresholds were performed at room temperature with a base pressure lower than $5\times10^{-9}$ Torr. During the absorption measurements, the CoO(100) single crystal was oriented so that the photons were incident at an angle of about 90$^o$ with respect to the sample surface. In order to minimize self-absorption effects [17], the angle of incidence was about 25$^o$ during the emission measurements. The emitted photons were always recorded at an angle, perpendicular to the direction of the incident photons, with the polarization vector parallel to the horizontal scattering plane. The counting rates were 3-6 counts/sec and the acquisition times 2-3 hours/spectrum, depending on the photon energy.

## 3 Calculational Details

The Co 3$d\rightarrow$2$p$ RIXS spectra of CoO were calculated as a coherent second-order optical process including interference effects using the Kramers-Heisenberg formula[18]:

$$I(\Omega,\omega)= \sum_f \left| \sum_i \frac{\langle f|D_q|i\rangle \langle i|D_q|g\rangle}{E_g+\Omega-E_i-i\Gamma_i/2} \right|^2$$

$$\times \delta(E_g+\Omega-E_f-\omega).$$

The $\Omega$ and $\omega$ denotes the excitation and emission energies, the $|g\rangle$, $|i\rangle$ and $|f\rangle$ are the ground, intermediate and final states with energies $E_g$, $E_i$ and $E_f$, $\hat{D}$ is the dipole operator and $\Gamma_i$ is the full width half maximum (FWHM) of the Lorenzian of each intermediate state representing the lifetime broadening which interfere between the different intermediate states. The values of the $\Gamma_i$s used in the calculations were 0.5 eV and 0.7 eV for the $L_3$ and $L_2$ thresholds, respectively [19]. The Slater integrals, describing $3d-3d$ and $3d-2p$ Coulomb and (super)exchange interactions, and spin-orbit constants were obtained by the Hartree-Fock method [20]. The effect of the configurational dependent hybridization was taken into account by scaling the Slater integrals to $F^k(3d3d)$ 80%, $F^k(2p3d)$ 80% and $G^k(2p3d)$ 80%. The ground state of the $Co^{2+}$ ion has $^4T_1$ character in $O_h$ symmetry. In order to take into account the polarization dependence and (super)exchange interactions, the calculations were made in the $C_{4h}$ basis set at 0 K. Two configurations were considered: $3d^7$ and $3d^8\underline{L}^{-1}$ for the initial and final states, and $2p^53d^8$, $2p^53d^9\underline{L}^{-1}$ for the intermediate states. The weights of the $3d^7$ and $3d^8\underline{L}^{-1}$ configurations in the ground state were 84% and 16%, respectively. The contribution of the $3d^9\underline{L}^{-2}$ configuration in the ground state was neglected here since its weight was estimated to be only in the order of ~ 1% in prior studies [21].
The single impurity Anderson model (SIAM) [22] with full multiplet effects was applied to describe the system. The crystal field and (super)exchange-interactions were taken into account by using a code by Butler [23] and the charge-transfer effect was implemented with a code by Thole and Ogasawara [24]. The SIAM parameters were chosen to reproduce the experiment as follows: the charge-transfer energy $\Delta$, defined as the energy difference between the center-of-gravity between the $3d^7$ and the $3d^8\underline{L}^{-1}$ configurations was 4.0 eV, the crystal-field splitting $10Dq$ was set to 0.5 eV and the (super)exchange field was applied in the direction of the polarization vector of the incoming photons. The scattering angle between the incoming and outgoing photons was fixed to $90^o$ and the calculations were made for the same geometry as the experimental one. The polarization dependence can be understood from a group theoretical consideration where according to the Wigner-Eckart theorem, the transition matrix elements are described by the Clebsch-Gordan coefficients in the $C_{4h}$ symmetry. The shape of the oxygen valence band, was appoximated by a function describing a circle with a width of 4.0 eV. The hybridization strength between the Co $3d$ band and the O $2p$ band where $V_{e_g}$ represents the hopping for the $Co^{2+}$ $e_g$ orbitals was taken to be 2.2 eV and 1.8 eV, for the ground and intermediate states, respectively. The smaller value of the hybridization strength of the intermediate states is due to the configurational dependence [25,26]. The value of the hybridization strength for the $Co^{2+}$ states of $t_{2g}$ symmetry $V_{t_{2g}}$ was taken as half of the value for the $e_g$ states $V_{e_g}$ which has previously been shown to be a reasonable empirical relation [27]. The (super)exchange interactions which correspond to strong effective magnetic fields were taken into account in the model Hamiltonian using a mean-field theory [28]. The parameters used in the calculations are summarized in Table I.

# 4 Results and Discussion

Figure 1 shows a set of RIXS spectra of CoO recorded at different excitation energies at the Co $2p_{3/2,1/2}$ thresholds. At the top, an XAS spectrum is shown (dots) where the excitation energies for the RIXS spectra are indicated by the arrows aimed at the main peaks and satellite structures. A calculated isotropic XAS spectrum with the corresponding multiplets including all symmetries is also included (full curve). The final states of the CoO XAS spectrum are the same as the intermediate states in the RIXS process and are well understood in terms of atomic transitions in a crystal field [21,29,30]. The main peaks in the XAS spectrum around ∼ 779.3 eV and ∼ 795.0 eV which are separated by the 2*p* spin-orbit splitting are made up of the crystal-field splitted $2p^5 3d^8$ configuration, hybridized with the $2p^5 3d^9 \underline{L}^{-1}$ manifold.

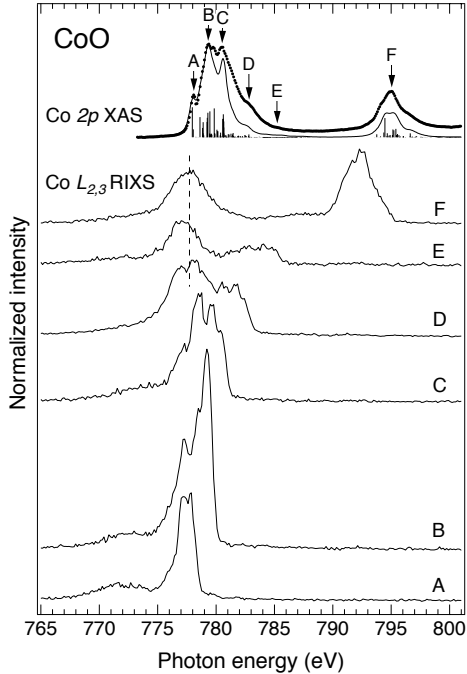

**Figure 1:** At the top is an XAS spectrum (dots) of CoO measured at the Co 2*p* edges compared to a calculated XAS spectrum (full curve) broadened with a Lorenzian (0.5 eV and 0.7 eV FWHM for the $2p_{3/2}$ and $2p_{1/2}$ thresholds, respectively) and a Gaussian profile of 0.35 eV. Below are measured $L_{2,3}$ RIXS spectra (full lines) on a photon energy scale excited at 778.1, 779.3, 780.5, 782.7, 785.2, 795.0 eV, denoted by the letters A-F, respectively.

The RIXS spectra in the lower part of Fig. 1 are plotted on an emission photon energy scale, normalized to the incoming photon flux, and were measured at the excitation energies denoted by the letters A-F from 778.1 eV up to 795.0 eV. As observed in Fig. 1, the spectral shape strongly depend on the excitation energy and show strong resonant enhancements at the $2p_{3/2}$ (spectrum B) and $2p_{1/2}$ (spectrum F) thresholds. The fluorescenece data basically contain peak structures of three different categories: recombination due to elastic $2p^5 3d^8 \rightarrow 3d^7$ transitions back to the ground state, also known as Rayleigh scattering, the resonating loss structures due to *d−d* excitations and charge-transfer excitations of the Raman scattering, and normal X-ray emission lines. In order to minimize the elastic contribution at threshold excitation, the emitted photons were recorded at an angle, perpendicular to the direction of the incident photons, with the polarization vector parallel to the horizontal scattering plane. The elastic and Raman scattering contributions disperse on the photon energy scale while other structures due to normal emission do not reveal any major energy shifts at all. In spectrum F, excited at the $L_2$ threshold, the main $L_3$ emission peak at ∼ 778 eV is due to normal X-ray fluorescence at constant photon energy as indicated by the dashed vertical line.

**Table 1:** The parameter values used in the Anderson impurity model calculations. κ is the scaling factor for the Slater integrals, Δ is the energy difference between the gravity centers of the $3d^7$ and the $3d^8 \underline{L}^{-1}$ configurations, $V_{eg1}$ and $V_{eg2}$ are the hybridization strengths for the eg orbitals in the ground and core-excited states, respectively. *W* is the O 2*p* bandwidth, *Q* is the core-hole potential, *U* is the on-site Coulomb interaction between the localized 3*d* electrons and 10*Dq* is the crystal-field splitting. The (super)exchange field was applied along the z-axis. All values, except for κ are in units of eV:s.

| κ | Δ | $V_{eg_1}$ | $V_{eg_2}$ | W | Q−U | 10Dq | Ex.field |
|---|---|---|---|---|---|---|---|
| 0.8 | 4.0 | 2.2 | 1.8 | 4.0 | 0.0 | 0.5 | 0.3 |

When the excitation energy is tuned to the charge-transfer satellite region in the absorption spectrum at D and E, it gives rise to intense emission lines at ~ 777 eV. The energy position of these lines is about ~ 1.0 eV lower than for the normal emission line. This distinct energy shift shows that the origin of this line is not normal fluorescence but rather due to charge-transfer as will be more discussed in detail below.

Figure 2 shows the $L_{2,3}$ RIXS data (dots) together with the results of SIAM calculations plotted as a function of the energy loss or Raman shift. The energy loss is derived from the RIXS spectra by subtracting the incident photon energy from the energy of the emitted photons. In order to enhance the spectral shape modifications, the spectra are normalized to the same peak heights. The calculations were made in the same geometry as the experimental one (see section II). The letters A-F denote the same excitation energies as in Fig. 1. As observed, the Raman scattering calculations are generally in good agreement with the experimental data although contributions from normal fluorescence are not included in the theory. The peak structure at 0 eV energy loss corresponds to the elastic recombination peak back to the $^4T_{1g}$-derived ground state. Other prominent features corresponding to Raman scattering appears at ~ 0.9 eV and ~ 2.0 eV. In addition, in the energy region 5-9 eV, there is a broad satellite structure with relatively low intensity. The Raman scattering peaks at ~ 0.9 eV and ~ 2.0 eV which stay at constant energy loss, identified in all spectra with different intensities, are essentially due to the ligand field splitting of the $3d^7$ final states. In Table II, the assignments of the loss structures are compared to those of absorption data [4,5] and EELS [6,7,8]. The lowest loss structure at ~ 0.9 eV above the ground state, most clearly observed in spectra A and C and as a shoulder in spectrum B, is due to the $^4T_{1g} \rightarrow ^4T_{2g}$ quartet-quartet transitions. The assignment of the ~ 2.0 eV loss structure in spectra B and C, also observed as a shoulder in spectrum A, is more difficult. We assign this feature to

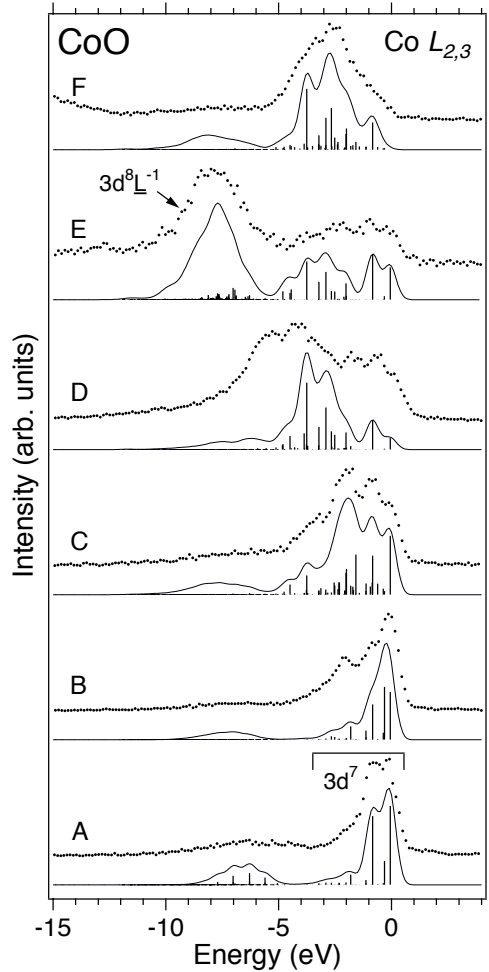

**Figure 2:** Calculated RIXS spectra (full lines) of CoO compared to the experimental spectra in Fig. 1 (dots) normalized to the same peak heights. The letters corresponds to the same excitation energies as in Fig. 1.

$^4T_{1g} \rightarrow {}^2T_{1g,2g}$ quartet-doublet spin-flip transition in agreement with EELS meaurements [6]. However, another EELS publication assign the ~ 2.0 eV loss feature to the $^4T_{1g} \rightarrow {}^4A_{2g}$ quartet-quartet transition [7].

In spectrum C, the shoulder at ~ 2.3 eV and the peak at ~ 3.2 eV are probably due to the $^4T_{1g} \rightarrow {}^4T_{1g}$ and $^4T_{1g} \rightarrow {}^4A_{2g}$ quartet-quartet transitions, respectively. However, the assignments are difficult and several calculations of measured energy-loss peaks in the energy region ~ 2.3-3.2 eV differ strongly for CoO and are sometimes also referred to as quartet-doublet spin-flip transitions [8]. The calculations of the spectral RIXS profiles reveal the sensitivity to the (super)exchange field which gives rise to a spectral weight transfer towards lower loss energies as it is increased. Thus, the magnitude of the (super)exchange energy of 0.3 eV needed to reproduce the experimental spectra gives important information about the degree of covalent bonding which gives rise to the antiferromagnetic alignment of the ions in the crystal field. The intense dispersing lines with loss energies of ~ 5.3 eV and ~ 8.0 eV in spectra D and E are reproduced by the model calculations as resonances of ligand $2p \rightarrow$ metal $3d$ charge-transfer excitations to $3d^8\underline{L}^{-1}$ final states. A comparison between Figures 1 and 2 shows that these final states remain at constant photon energy on the emission energy scale and disperse on the energy loss scale. Similar resonantly enhanced energy loss structures as those observed in spectra D-E have been assigned to have charge-transfer origin in RIXS spectra of rare earths compounds [31].

**Table 2:** Ground state and final state energies of $d-d$ transitions in CoO(100) measured by RIXS, EELS and optical absorption. All values are in eV:s.

| Symmetry | RIXS | EELS | Opt. abs. |
|---|---|---|---|
| $^4T_{1g}$ | 0 | 0 | 0 |
| $^4T_{2g}$ | 0.9 | 0.85 | 0.9-1.04 |
| $^2T_{1g}$ | 2.0 | 2.05 | 2.04 |
| $^2T_{2g}$ | 2.0 | 2.05 | 2.06 |
| $^4T_{1g}$ | 2.3 | 2.25 | 2.3 |
| $^4A_{2g}$ | 3.2 | 3.2 | 2.15 |

The relative peak positions of the low-energy loss structures in RIXS, EELS and optical absorption spectra are different from the results of photoemission measurements [2]. The charge-transfer nature of CoO results in spectral features due to both $3d^6$ and $3d^7\underline{L}^{-1}$ final states to be observed in valence-band photoemission. Peak structures at ~ 1.7 eV and ~ 3.8 eV have been interpreted to have $3d^7\underline{L}^{-1}$ final state character, while $3d^6$ final states are observed at ~ 10.4 eV binding energy [9,13]. A broader double structure in the region ~ 5.1-7.6 eV is due to the O $2p$ band. The $3d^7\underline{L}^{-1}$ final state in RPES is believed to originate in a process in which the initial $3d^7$ ground state is ionized to a $3d^6$ state which is then screened by the charge-transfer from the ligand to the metal ion. Thus, the ligand-field splitted $3d^7\underline{L}^{-1}$ final states in RPES appears at relatively higher binding energies than the loss structures in RIXS (see Table II) and represents excited states not directly reflecting the electronic structure of the ground state. The RIXS technique is here shown to be very sensitive for detecting quartet-quartet, quartet-doublet $d-d$ excitations as well as the important $d^7 \to d^8$ charge-transfer excitations in CoO. Note that the $d^8$ final states observed in the RIXS spectra cannot be observed in RPES but instead appears at ∼ 4.0 eV in inverse photoemission[32]. The difference in the energy positions of the peaks between the different spectroscopical techniques is a result of probing the systems with different final states in comparison to the ground state. Since the excitation-deexcitation process is charge-neutral, RIXS is generally a very powerful tool for investigating the extended multiplet structure of the ground and low-energy excited configurations of transition metal compounds dominated by charge-fluctuations which is not fully accessible with other spectroscopical techniques.

# 5  Summary

The electronic structure of CoO has been probed at the Co $2p$ absorption thresholds by resonant inelastic soft-X-ray scattering. By changing the incoming photon energy, the contribution from elastic and Raman scattering is distinguished from charge-transfer excitations and normal fluorescence. Raman scattering due to quartet-quartet, quartet-doublet $d-d$ excitations in the crystal field are identified. Pronounced energy loss structures which disperse below the elastic peak are identified as due to charge-transfer excitations to the $3d^8\underline{L}^{-1}$ final state. Anderson impurity model calculations are consistent with the present experimental findings implying high sensitivity to the crystal-field and (super)exchange interactions.

# 6  Acknowledgments

This work was supported by the Swedish Natural Science Research Council (NFR) and the Göran Gustafsson Foundation for Research in Natural Sciences and Medicine.